\documentclass[aps,pre,reprint,showpacs,amsmath,amssymb,floatfix,10pt]{revtex4-1} % PRL -superscriptaddress

\usepackage[pdftex]{graphicx}       % Include figure files
\usepackage{txfonts}        % Times New Roman text & math font
\usepackage{amsmath}
\usepackage{amssymb}
\usepackage{mathtools}
\usepackage{mathrsfs}	% Extra script font
\newcommand{\Tcn}{T_\text{a}^{(n)}}
\newcommand{\Tcinf}{T_\text{a}}

\newcommand{\eref}[1]{Eq.~\eqref{#1}}
\newcommand{\fref}[1]{Fig.~\ref{#1}}
\newcommand{\sref}[1]{Section \ref{#1}}
\newcommand{\tref}[1]{Table \ref{#1}}

\usepackage{hyperref} % this line at end of preamble
\hypersetup{colorlinks,citecolor=blue,filecolor=blue,linkcolor=blue,urlcolor=blue}

\begin{document}

\title{Adsorption of neighbor-avoiding walks on the simple cubic lattice}

\author{C. J. Bradly} \email{chris.bradly@unimelb.edu.au}
\author{A. L. Owczarek}\email{owczarek@unimelb.edu.au}
\affiliation{School of Mathematics and Statistics, University of Melbourne, Victoria 3010, Australia}
\author{T. Prellberg} \email{t.prellberg@qmul.ac.uk}
\affiliation{School of Mathematical Sciences, Queen Mary University of
  London, Mile End Road, London, E1 4NS, United Kingdom}
\date{\today}

\begin{abstract}
We investigate neighbor-avoiding walks on the simple cubic lattice in the presence of an adsorbing surface.
This class of lattice paths has been less studied using Monte Carlo simulations. Our investigation follows on from our previous results using self-avoiding walks and self-avoiding trails.
The connection is that neighbor-avoiding walks are equivalent to the infinitely repulsive limit of self-avoiding walks with monomer-monomer interactions. Such repulsive interactions can be seen to enhance the excluded volume effect.
We calculate the critical behavior of the adsorption transition for neighbor-avoiding walks, finding a critical temperature \smash{$\Tcinf=3.274(9)$} and a crossover exponent \smash{$\phi=0.482(13)$}, which is consistent with the exponent for self-avoiding walks and trails, leading to an overall combined estimate for three dimensions of \smash{$\phi_\text{3D}=0.484(7)$}. 
While questions of universality have previously been raised regarding the value of adsorption exponents in three dimensions, our results indicate that the value of $\phi$ in the strongly repulsive regime does not differ from its non-interacting value. However, it is clearly different from the mean-field value of $1/2$ and therefore not super-universal.

\end{abstract}

\maketitle

%%%%%%%%%%%%%%%%%%%%%%%%%%%%%%%%%%%%%%%%%%%%%%%%%%%%%%%%%%%%%%
%%%%%%%%%%%%%%%%%%%%%%%%%%%%%%%%%%%%%%%%%%%%%%%%%%%%%%%%%%%%%%
\section{Introduction}
\label{sec:Intro}

The critical phenomenon associated with the adsorption of polymers in dilute solution onto a surface is a widely studied problem in statistical physics \cite{Eisenriegler1982,DeBell1993,Vrbova1996,Vrbova1998,Vrbova1999,Grassberger2005,Owczarek2007,Luo2008,Klushin2013,Plascak2017}. 
This topic has applications to interfacial phenomena such as adhesion and general applications in biology \cite{Milner1991,Diaz2007,Gennes1980,Meredith1998}.
In the thermodynamic limit of infinitely long polymers, the adsorbed fraction of the polymer $u_\infty$ is zero at high temperatures where the configuration of the polymer is dominated by entropic repulsion, forming an expanded phase where the polymer is desorbed from the surface.
If there is an attractive surface-monomer interaction, then below some temperature $\Tcinf$ there is an adsorbed phase where $u_\infty$ is positive.
This transition is continuous and the polymer ensemble displays a critical phenomenon \cite{DeBell1993}. 
For finite lengths, the scaling of the adsorbed fraction is determined by a critical exponent $\phi$: 
\begin{equation}
	u_n\sim n^{\phi-1},
	\label{eq:OrderParameter}
\end{equation}
where $\phi$ takes on a non-integer value when $T=T_a$.
This exponent was initially considered to have the same value in all dimensions making it super-universal.
This hypothesis was supported by results for two dimensions where $\phi=1/2$, matching the mean-field value predicted for all dimensions above the upper critical dimension of 4 as well as early results for three dimensions \cite{Hegger1994,Metzger2003}.
More recently, numerical simulations have found that \smash{$\phi\neq 1/2$} for three dimensions.
Our own estimate \smash{$\phi=0.484(4)$} from a study of adsorbing self-avoiding walks and trails \cite{Bradly2018} is in agreement with other recent Monte Carlo studies finding \smash{$\phi=0.484(3)$} \cite{Grassberger2005}, \smash{$\phi=0.492(4)$} \cite{Plascak2017} and \smash{$\phi=0.483(3)$} \cite{Klushin2013}.
On the other hand, others have found values of $\phi$ that exceed $1/2$ \cite{Luo2008,Taylor2009}.

These results assume a good solvent so that away from the surface the polymers have an extended configuration due to the excluded volume effect, making self-avoiding walks (SAWs) on a lattice the canonical model.
The effect of solvent quality is usually modeled by adding a monomer-monomer interaction to the self-avoiding walk and varying the interaction strength.
Recently, it has been suggested that $\phi$ may not be universal as a result of this interaction \cite{Luo2008}.
Plascak {\em et al.} \cite{Plascak2017} considered a combined model of adsorbing SAWs with varying strength of monomer-monomer interactions. 
They found that $\phi$ changes value as the monomer-monomer interaction strength is changed. 
In the strongly repulsive regime they found a small but significant reduction in the critical exponent $\phi$ compared to the non-interacting value.
In this paper we therefore consider the strongly repulsive regime as any change in this exponent due to repulsive interactions should be most apparent in this limit.

The canonical representation of a polymer in dilute solution is a self-avoiding path on a lattice.
In the context of lattice paths, altering the excluded volume effect can also be modeled by different classes of paths.
The standard model of self-avoiding walks (SAWs) on a lattice are a subset of self-avoiding trails (SATs) which have the weaker restriction that bonds may not overlap but lattice sites may be occupied by multiple steps.
Similarly, neighbor-avoiding walks (NAWs) are a subset of SAWs with the stronger restriction that non-consecutive steps in the walk cannot be adjacent.
Compared to SATs and SAWs, NAWs have received only some attention \cite{Jensen1998,Bennett-Wood1998} and this has not included monomer-monomer or monomer-surface interactions.
%The connection with monomer-monomer interactions is strongly repulsive interactions for say, SAWs, will prevent adjacent sites from being occupied.
%Similarly, SAWs are the repulsive limit of interacting SATs.
Neighbor-avoiding walks and self-avoiding walks can be considered as the infinitely repulsive limit of interacting self-avoiding walks and interacting self-avoiding trails, respectively.
Instead of attempting to simulate a combined model with adsorption and strong repulsive interactions, NAWs can be simulated directly.

In this work we combine new data for the adsorption on to a surface of neighbor-avoiding walks on the simple cubic lattice with our previous results for self-avoiding walks and self-avoiding trails.
We find no evidence that the critical exponent $\phi$ is non-universal in these limits. We also confirm that it is indistinguishable from the $1/\delta$ exponent controlling the shift of the critical temperature for finite lengths.

%%%%%%%%%%%%%%%%%%%%%%%%%%%%%%%%%%%%%%%%%%%%%%%%%%%%%%%%%%%%%%
%%%%%%%%%%%%%%%%%%%%%%%%%%%%%%%%%%%%%%%%%%%%%%%%%%%%%%%%%%%%%%
\section{Lattice models}
\label{sec:Models}

To model adsorption we consider lattice paths $\psi_n$ of length $n$ restricted to lie on one side of an impermeable surface defined by \smash{$z=0$} for the simple cubic lattice. 
Apart from one end that is fixed at the origin, each of the $m$ contacts with the surface contributes an energy $\epsilon_\text{a}$ and corresponding Boltzmann weight $\kappa^m$, where \smash{$\kappa = \exp(\epsilon_\text{a}/k_\text{B}T)$}. 
This gives the partition function for lattice paths of length $n$
\begin{equation}
    Z_n(\kappa) = \sum_{\psi_n} \kappa^m,
    \label{eq:Partition}
\end{equation}
from which we calculate the internal energy
\begin{equation}
    u_n (\kappa) = \frac{\langle m \rangle}{n},
    \label{eq:InternalEnergy}
\end{equation}
which serves as our order parameter. 

%%%%%%%%%%%%%%%%%%%%%%%%%%%%%%%%%%%%%%%%%%%%%%%%%%%%%%%%%%%%%%
\begin{figure}[t!]
	\includegraphics[width=\columnwidth]{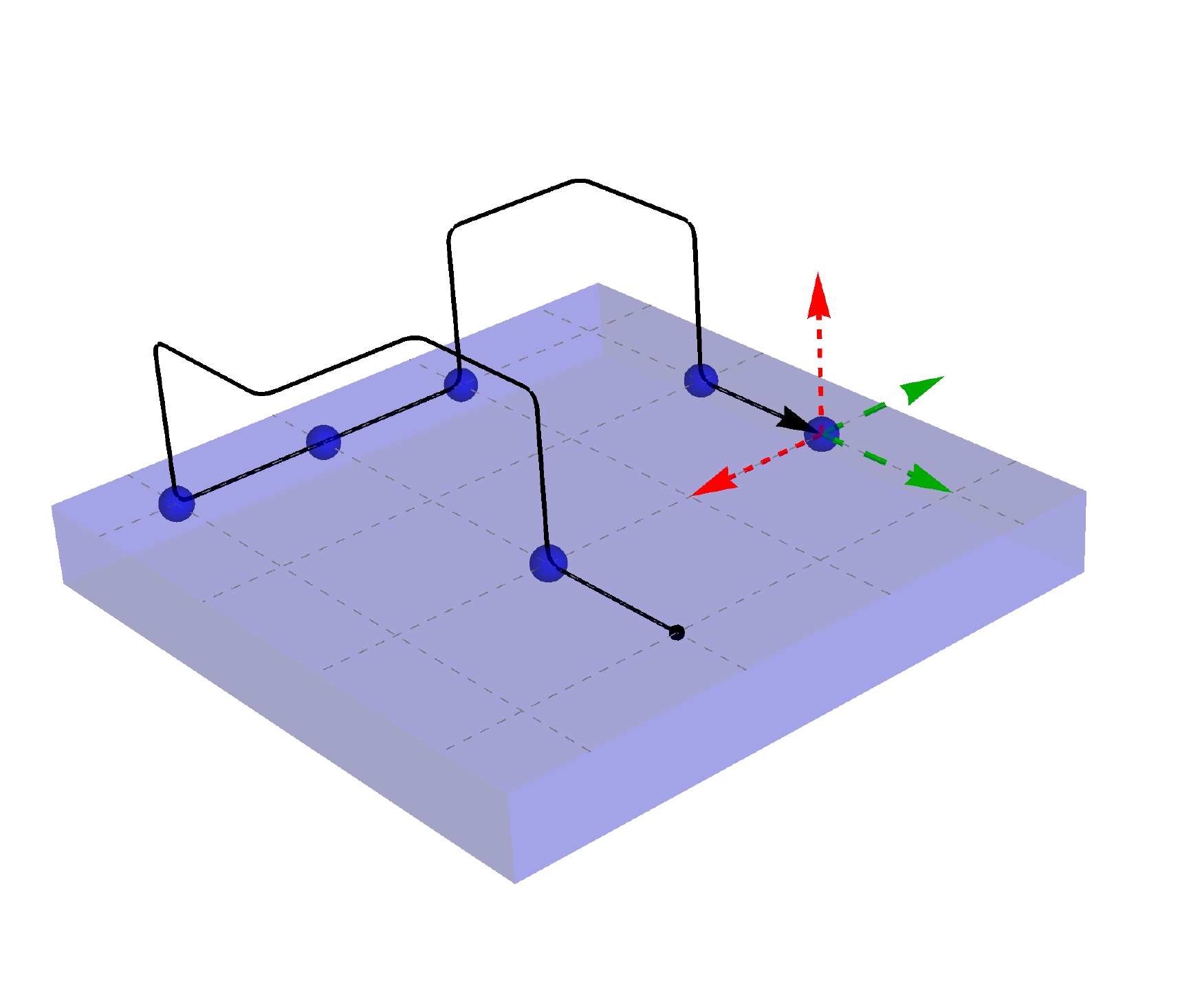}%
	\caption{A neighbor-avoiding walk on the simple cubic lattice near a surface. Green dashed arrows and red dotted arrows mark valid and invalid next steps, respectively. Blue circles mark monomers that are interacting with the surface.}%
	\label{fig:NAWValidSteps}%
\end{figure}
%%%%%%%%%%%%%%%%%%%%%%%%%%%%%%%%%%%%%%%%%%%%%%%%%%%%%%%%%%%%%%

%%%%%%%%%%%%%%%%%%%%%%%%%%%%%%%%%%%%%%%%%%%%%%%%%%%%%%%%%%%%%%
\begin{figure}[t!]
	\includegraphics[width=\columnwidth]{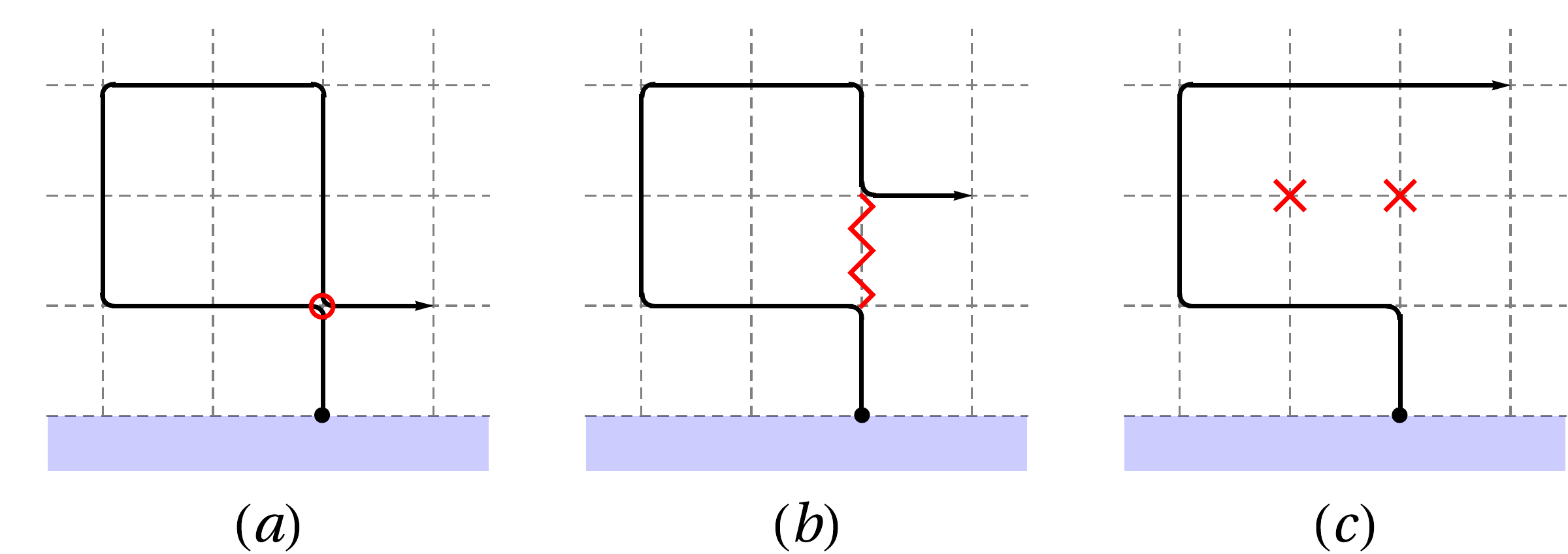}%
	\caption{Comparison on the square lattice of (a) an interacting self-avoiding trail (ISAT) with onsite contact interaction, (b) an interacting self-avoiding walk (ISAW) with near-neighbor interaction and (c) a neighbor-avoiding walk where neighboring points are avoided. 
	In each case, the model to the right can be viewed as the infinitely repulsive limit of the monomer-monomer interaction.}
	\label{fig:WalkComparisons}%
\end{figure}
%%%%%%%%%%%%%%%%%%%%%%%%%%%%%%%%%%%%%%%%%%%%%%%%%%%%%%%%%%%%%%

Solvent quality of polymers is modeled by monomer-monomer interactions between non-consecutive lattice sites that are adjacent in some way, depending on the lattice model. 
For interacting self-avoiding walks (ISAW) each lattice site interacts with its adjacent neighbors, while for interacting self-avoiding trails (ISAT) the interaction is found at multiply-visited lattice sites, see \fref{fig:WalkComparisons}(a) and (b). 
In either case, each of the $b$ pair-wise interactions contributes an energy $-\epsilon_\text{mm}$ for  and is thus weighted by a factor $\omega^b$ where \smash{$\omega=\exp(\epsilon_\text{mm}/k_\text{B}T)$}.
There is a critical temperature $\omega_c>0$ where the polymer undergoes collapse from an extended coil to a globule.
This model has been extensively used to study the $\theta$-point collapse of real polymers.
We note however, that while both ISAW and ISAT have been used extensively to model $\theta$-point collapse of real polymers, they are believed to be in different universality classes \cite{Shapir1984,Lim1988,Owczarek1995,Prellberg1995}.

A combined model of interacting paths near an adsorbing surface, has partition function 
\begin{equation}
Z_n(\kappa,\omega) = \sum_{\psi_n} \kappa^m\omega^b.
\label{eq:CombinedPartition}
\end{equation}
Recently, Plascak {\em et al.}~\cite{Plascak2017} investigated this case and found that the critical exponent $\phi$ associated with the adsorption transition varied with the ratio of the interaction energies \smash{$\epsilon_\text{mm}/\epsilon_\text{a}$}.
In particular, they consider that the surface-monomer interaction is always attractive so as to allow adsorption but the monomer-monomer interaction can vary in strength from attractive through non-interacting and even to strongly repulsive.

The limit of increasingly repulsive monomer-monomer interactions, \smash{$\epsilon_\text{mm}\to-\infty$}, is equivalent to \smash{$\omega\to 0$} for fixed $\kappa$.
%Numerical simulations that approximate the partition function \eref{eq:CombinedPartition} cannot calculate this limit directly except by extrapolating successively smaller values of $\omega$.
%But of course this limit exists since self-avoiding walks are really modeling the excluded volume effect of real polymers and so strong monomer-monomer repulsion is really just an enhancement of the excluded volume effect.
This is modeled directly by altering the class of lattice paths we are considering, accomplished by a simple change to the rules of the lattice path.
So, rather than adjacent lattice sites in ISAW counting towards the number of monomer-monomer interactions, the walk is prevented from even occupying a site that is already adjacent to another occupied site, if the repulsion is strong enough.
We call this subset of SAWs neighbor-avoiding walks (NAWs), and an example is shown in \fref{fig:NAWValidSteps}.
Similarly, in the strongly repulsive limit of an ISAT model, the monomer-monomer repulsion prevents multiply-visited sites, and the trail simply becomes a non-interacting SAW.
These comparisons are illustrated in \fref{fig:WalkComparisons}.

Specifically, the partition function for NAWs is the same as the strongly repulsive limit of the partition function of ISAW.
The same is true for SAWs, in that their partition function is the same as the strongly repulsive limit of the partition function of ISAT.
That is
\begin{alignat}{1}
	Z_n^{(\text{NAW})}(\kappa) &= Z_n^{(\text{ISAW})}(\kappa,\omega=0)\quad\mbox{and} \\
	Z_n^{(\text{SAW})}(\kappa) &= Z_n^{(\text{ISAT})}(\kappa,\omega=0),
	\label{eq:MatchedPartition}
\end{alignat}
where the superscript indicates the class of paths over which the sum in \eref{eq:Partition} or \eref{eq:CombinedPartition} is taken, and the interaction variable $\omega$ refers to on-site contact interactions for ISAT and to nearest-neighbor interactions for ISAW, respectively.
Clearly, the critical adsorption temperature is different for each class of lattice paths, and also depends on $\omega$ in the case of the interacting models.
%{\bf (there's probably a better way to express this)}

The great benefit of the restriction to specific non-interacting ensembles is that it is much easier to simulate the adsorption-only model for different classes of lattice paths than to do a combined interacting model, allowing for simulation of the adsorption transition for larger system sizes.
%From this point we can study the critical behaviour of the adsorption transition with better data from larger $n$ simulations.

%Lastly, we note that we cannot make a similar argument that applies to the infinitely attractive limit of monomer-monomer interactions.
%From the point of view of classes of lattice paths, we do not have a clear model for interacting NAWs, and viewing SATs as infinitely attractive SAWs ignores the fact that a collapsed SAW does not necessarily reduce to a SAT.
%Even if this regime could be expressed by altering the class of lattice models, the limit \smash{$\omega\to\infty$} means $\omega$ passes through the critical point of the collapse transition and the situation is not so clear.

%%%%%%%%%%%%%%%%%%%%%%%%%%%%%%%%%%%%%%%%%%%%%%%%%%%%%%%%%%%%%%
%%%%%%%%%%%%%%%%%%%%%%%%%%%%%%%%%%%%%%%%%%%%%%%%%%%%%%%%%%%%%%
\section{Scaling laws and critical exponents}
\label{sec:Scaling}

At the critical point for long chains the order parameter scales as $u_n\sim n^{\phi-1}$ but for finite $n$ it is necessary to also include finite-size scaling correction terms:
\begin{equation}
    u_n \sim n^{\phi-1} f_u^\text{(0)}(x) [1 + n^{-\Delta}f_u^\text{(1)}(x) +\ldots ],
    \label{eq:UnScaling}
\end{equation}
where the $f^{(i)}$ are  finite-size scaling functions of the scaling variable \smash{$x=(\Tcinf-T)\,n^{1/\delta}$} and \smash{$\Delta\lesssim 1$} is the first correction-to-scaling term. 
The exponent $1/\delta$ therefore describes the \emph{crossover} around the adsorption critical point. 
It can also be described as  the shift exponent associated with the deviation of temperature from the critical point. 
That is, the finite-length critical temperature differs from the infinite-length critical temperature according to
\begin{equation}
    \Tcn \sim \Tcinf + n^{-1/\delta} f_T^\text{(0)}(x) [1 + n^{-\Delta}  f_T^\text{(1)}(x)+\ldots].
    \label{eq:TempScaling}
\end{equation}
Conventional scaling arguments show that the exponents $\phi$ and $1/\delta$ are the same and one can be derived from the other \cite{Eisenriegler1982,Bradly2018}.
Recently, however, Luo \cite{Luo2008} conjectured that $\phi$ and $1/\delta$ may be different in three dimensions.
Under this assumption, other numerical work has been carried out to test this hypothesis, finding different numerical values for $\phi$ and $1/\delta$ on top of the dependence on monomer-monomer interaction strength \cite{Plascak2017,Martins2018}.

The first step is to extract $1/\delta$ directly from the log-derivative of $u_n$,
\begin{equation}
    \Gamma_n(\kappa) = \frac{d\log u_n}{dT} =
	(\log\kappa)^2
	\frac{\langle m^2 \rangle - \langle m \rangle^2}{\langle m \rangle}.
    \label{eq:LogDerivative}
\end{equation}
which is expected to have critical scaling form
%Considering the scaling ansatz of $u_n$ \eref{eq:UnScaling}, near the critical point the maximum of $\Gamma_n$ should scale as
\begin{equation}
    \max \Gamma_n \sim n^{1/\delta} f_\Gamma^\text{(0)}(x) [1 + n^{-\Delta} f_\Gamma^\text{(1)}(x) +\ldots].
    \label{eq:GammaScaling}
\end{equation}
The quantity $\Gamma_n$ is related to the specific heat, whose peaks are often used to locate second-order transitions.
However, this approach is known to be inaccurate for locating the adsorption transition \cite{Rensburg2004}. 
It is usually assumed that $x$ is small enough to use \eref{eq:GammaScaling} to determine $1/\delta$, but we will see that this is not generally a good approximation.

The main way to estimate $\phi$ is to determine the finite-size critical temperatures $\Tcn$, then use \eref{eq:TempScaling} and the value of $1/\delta$ to find $\Tcinf$. 
As before, the scaling variable $x$ is small and $\phi$ can be found by fitting the data to \eref{eq:UnScaling}.
The accuracy of this method depends on how we locate the $\Tcn$.
We use four methods of calculating $\Tcn$, which we list here for reference. 
For further details and comments on the accuracy of each method see reference \cite{Bradly2018}.

The first method, labeled ``$\Gamma$'', is to consider the locations of $\max\Gamma_n$, used to estimate $1/\delta$ from \eref{eq:GammaScaling}, as estimates of $\Tcn$.
Second, we calculate the Binder cumulant
\begin{equation}
    U_4(\kappa) = 1 - \frac{1}{3}\frac{\langle m^4 \rangle}{\langle m^2 \rangle^2},
    \label{eq:Binder}
\end{equation}
which tends toward a universal constant value at the critical point in the limit of large $n$ \cite{Binder1981} . 
Intersections of curves of $U_4$ at different $n$ with the curve at fixed \smash{$n_\text{min}=128$} are used to locate the finite-size critical temperatures. 
This method is labeled ``BC''.

The third method, labeled ``R2'', uses the mean-squared end-to-end radius. 
In the presence of a surface we distinguish between the transverse and perpendicular components, with respect to the surface.
For either component \smash{$i=\parallel,\perp$},
\begin{equation}
    \langle R_{i}^2\rangle_n \sim n^{2\nu_i},
    \label{eq:R2Scaling}
\end{equation}
and the Flory exponent $\nu_i$ depends on the phase and dimension of the system and is calculated by simply inverting \eref{eq:R2Scaling}:
\begin{equation}
   \nu_i = \frac{1}{2} \log_2\frac{\langle R_{i}^2\rangle_n}{\langle R_{i}^2\rangle_{n/2}}.
   \label{eq:NuRatio}
\end{equation}
In the desorbed phase both perpendicular and transverse components of $\langle R^2\rangle_n$ scale as per the $d$-dimensional bulk. 
In the adsorbed phase the components of $\nu$ differ, $\nu_\parallel$ takes on the {(d-1)}-dimensional bulk value and $\nu_\perp$ vanishes.
At intermediate temperatures the components of $\nu$ cross and in fact the intersections locate the finite-size critical temperatures $\Tcn$.

%%%%%%%%%%%%%%%%%%%%%%%%%%%%%%%%%%%%%%%%%%%%%%%%%%%%%%%%%%%%%%
\begin{figure*}[t!]
	\centering
	\includegraphics[width=\textwidth,trim={0 1.2cm 0 1.2cm},clip]{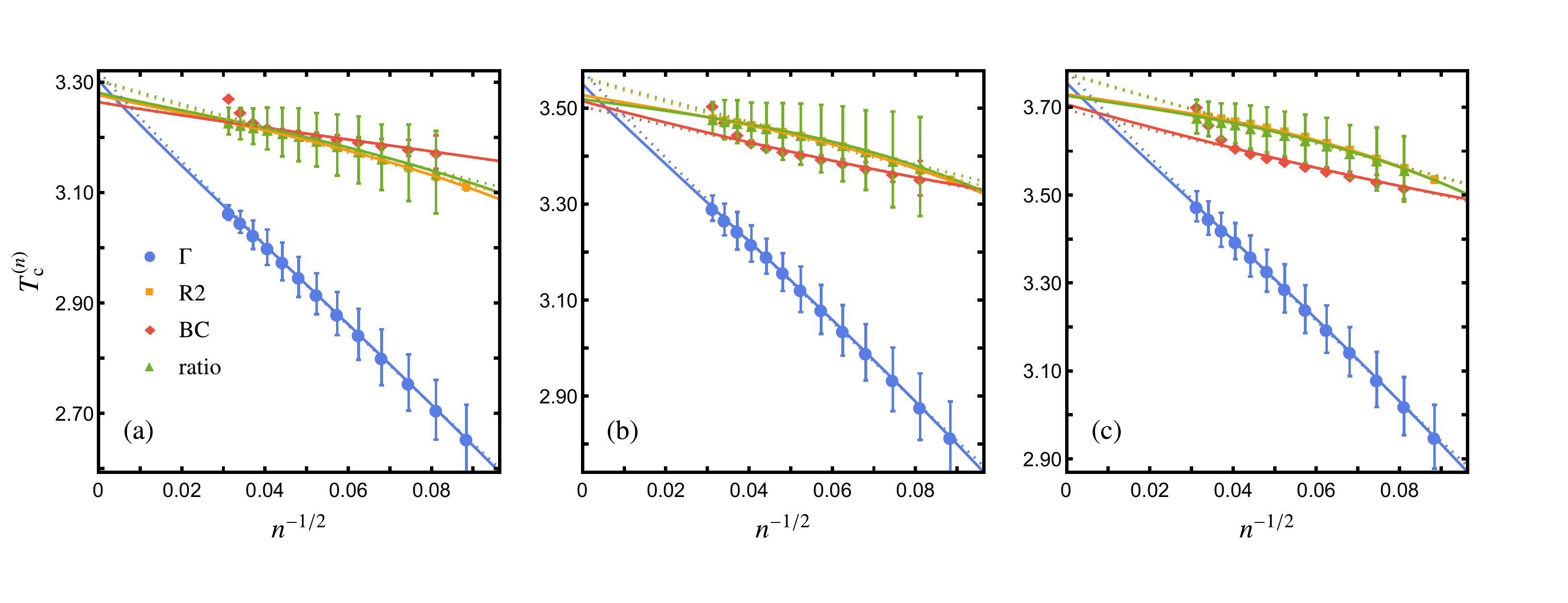}
	\caption{Finite-size critical temperatures for three-dimensional lattice models (a) NAWs, (b) SAWs and (c) SATs on the simple cubic lattice. For each of the four methods the solid lines are fits with correction to scaling and dotted lines are power-law only. Data for SAW and SAT models are from \cite{Bradly2018}.}
	\label{fig:3DTemperatures}
	\vspace{-0.5cm}
\end{figure*}
%%%%%%%%%%%%%%%%%%%%%%%%%%%%%%%%%%%%%%%%%%%%%%%%%%%%%%%%%%%%%%

The fourth method, labeled ``ratio'', is to calculate the exponent $\phi$ from the leading term of the order parameter. 
That is, we invert \eref{eq:OrderParameter} to obtain 
\begin{equation}
   \phi = 1 + \log_2\frac{u_{n}}{u_{n/2}},
   \label{eq:PhiRatio}
\end{equation}
which is evaluated over a range of $n$. 
As a function of temperature, $\phi$ vanishes at high temperatures and tends to unity far below the critical temperature.
Similar to the ``R2'' method, curves of \eref{eq:PhiRatio} cross over between these regimes and intersect near the critical temperature.
The position of these intersections for different values of $n$ are estimations of $\Tcn$.

As well as locating $\Tcn$ for use in the finite-size scaling ansatz of \eref{eq:UnScaling}, we can view \eref{eq:PhiRatio} as a ``direct'' estimate of $\phi$ over a range of finite $n$. 
In the limit \smash{$n\to\infty$}, these values extrapolate to an alternative estimate of $\phi$ without reference to the scaling form \eref{eq:UnScaling} and its dependence on locating the critical temperatures.

%%%%%%%%%%%%%%%%%%%%%%%%%%%%%%%%%%%%%%%%%%%%%%%%%%%%%%%%%%%%%%
%%%%%%%%%%%%%%%%%%%%%%%%%%%%%%%%%%%%%%%%%%%%%%%%%%%%%%%%%%%%%%
\section{Results}
\label{sec:Results}

%%%%%%%%%%%%%%%%%%%%%%%%%%%%%%%%%%%%%%%%%%%%%%%%%%%%%%%%%%%%%%
%%%%%%%%%%%%%%%%%%%%%%%%%%%%%%%%%%%%%%%%%%%%%%%%%%%%%%%%%%%%%%
%\subsection{Numerical simulation}
%\label{sec:Numerical}
We simulated SATs, SAWs and NAWs using the flatPERM algorithm \cite{Prellberg2004}, an extension of the pruned and enriched Rosenbluth method (PERM) \cite{Grassberger1997}. 
FlatPERM is a chain-growth algorithm so at each step in the simulation a lattice path of length $n$ is grown to length $n+1$ by choosing an available site adjacent to the current endpoint.
The ruleset for what qualifies as an available site determines the class of lattice model that is grown.
SATs have the restriction that bonds between sites may not overlap, although individual lattice sites may be multiply occupied.
SAWs have the additional restriction that each lattice site may be occupied at most once.
Finally, NAWs are like SAWs but with the extra restriction that a valid next step must not only be unoccupied but it must also have no adjacent occupied sites, refer to \fref{fig:NAWValidSteps}.
This is achieved by checking the neighbors of each point in the atmosphere of the endpoint of a walk and incurs very little extra computational cost.
In this work we used the flatPERM algorithm to simulate walks and trails on the simple cubic lattices up to length $1024$. 
We run 10 completely independent simulations for each case to estimate the statistical error of thermodynamic averages.
Details of the simulations run in this work are summarized in \tref{tab:SimDetails}. 

The main output of the simulation is the density of states $W_{n,m}$ of walks/trails of length $n$ with $m$ contacts with the surface, for all \smash{$n\le N_\text{max}$}. 
All thermodynamic quantities given in \sref{sec:Scaling} are then given by the weighted sum
\begin{equation}
    \langle Q \rangle(\kappa) = \frac{\sum_{m} Q_m\kappa^m W_{n,m}}{\sum_{m} \kappa^m W_{n,m}}.
    \label{eq:FPQuantity}
\end{equation}

%%%%%%%%%%%%%%%%%%%%%%%%%%%%%%%%%%%%%%%%%%%%%%%%%%%%%%%%%%%%%%
\setlength{\tabcolsep}{4pt}
\begin{table}[b!]
	\caption{Details of flatPERM simulations. In all cases the number of samples and effectively independent samples is the average of 10 independent runs. 
	%Simulations on the hexagonal lattice at fixed critical weight \smash{$\kcinf=1+\sqrt{2}$} are to longer length.
	}
	\begin{tabular}{lrrrr}
	\hline \hline
	& \multicolumn{1}{l}{Max} &  &	\multicolumn{1}{l}{Samples at} & \multicolumn{1}{l}{Ind.~samples} \\ [-1pt]
	Model & \multicolumn{1}{l}{length} & \multicolumn{1}{l}{Iterations }&  \multicolumn{1}{l}{max length}	& \multicolumn{1}{l}{max length}	\\ \hline
	NAW & 1024 & $4.4\times 10^5$ & $3.6\times 10^{10}$ & $5.1\times 10^8$ \\
	SAW & 1024 & $4.4\times 10^5$ & $3.5\times 10^{10}$ & $5.4\times 10^8$ \\
	SAT & 1024 & $4.4\times 10^5$ & $3.4 \times 10^{10}$ & $5.9\times 10^8$ \\ 
	\hline \hline
	\end{tabular}
	\label{tab:SimDetails}
\end{table}
%%%%%%%%%%%%%%%%%%%%%%%%%%%%%%%%%%%%%%%%%%%%%%%%%%%%%%%%%%%%%%

We now compare results for NAWs with our previous data for SAWs and SATs \cite{Bradly2018}.
First, we show in \fref{fig:3DTemperatures} values of $\Tcn$ for the four finite-size scaling methods and the fits to \eref{eq:TempScaling} to determine the critical temperature $\Tcinf$.
For NAWs the extrapolated values for each method are listed in \tref{tab:ExponentResults}. 
For specific values for SAWs and SATs see Table II in \cite{Bradly2018}.
For all lattice models and methods, the inclusion of the correction to scaling term is a significant improvement over a power-law only approach and is not dependent on the precise value of $\Delta$, provided that \smash{$0.5\lesssim\Delta\lesssim 1$}.
In the limit of long chains, all methods are in good agreement and we find an average value \smash{$\Tcinf=3.274(9)$} for NAWs.
This is less than the critical temperature for SAWs, \smash{$\Tcinf=3.520(6)$}, indicating that NAWs have an enhanced excluded volume effect and is in accordance with the estimate of Plascak {\em et al.} \cite{Plascak2017} for SAWs with large repulsive monomer-monomer interaction.
In turn, the critical temperature for SAWs is less than the critical temperature for SATs, \smash{$\Tcinf=3.720(12)$}.

The results for NAWs reaffirm a few issues with some of the methods of analysis.
In all cases it is clear that the peaks of $\Gamma_n$ are a poor way to locate the critical temperature at finite $n$ compared to the other methods.
That is, the assumption that the scaling variable $x$ is small breaks down for \eref{eq:GammaScaling}.
While the extrapolation to large $n$ matches the other methods, this raises the question of the validity of estimating the exponent $1/\delta$ from $\max\Gamma_n$.

Another issue is with the Binder cumulant method (red markers in \fref{fig:3DTemperatures}), where there is a strong dependence on the minimum value of $n$ used to find the intersections of \eref{eq:Binder}. 
This dependence accounts for the different location of the red curve relative to the other estimates for NAWs compared to SAWs and SATs in \fref{fig:3DTemperatures}.
 %and is not due to the lattice model themselves.
Also, the estimates from this method diverge at larger $n$, where small inaccuracies in the simulation weights are amplified when calculating the fourth-order moment $\langle m^4 \rangle$. 
For this method to be accurate at a given value of $n$ would require simulating walks to lengths larger than $n$.
%This is particularly true for NAWs where the enhanced swelling causes a greater underestimation of $U_4$.

\begin{figure}[t!]
	\centering
	\includegraphics[width=\columnwidth]{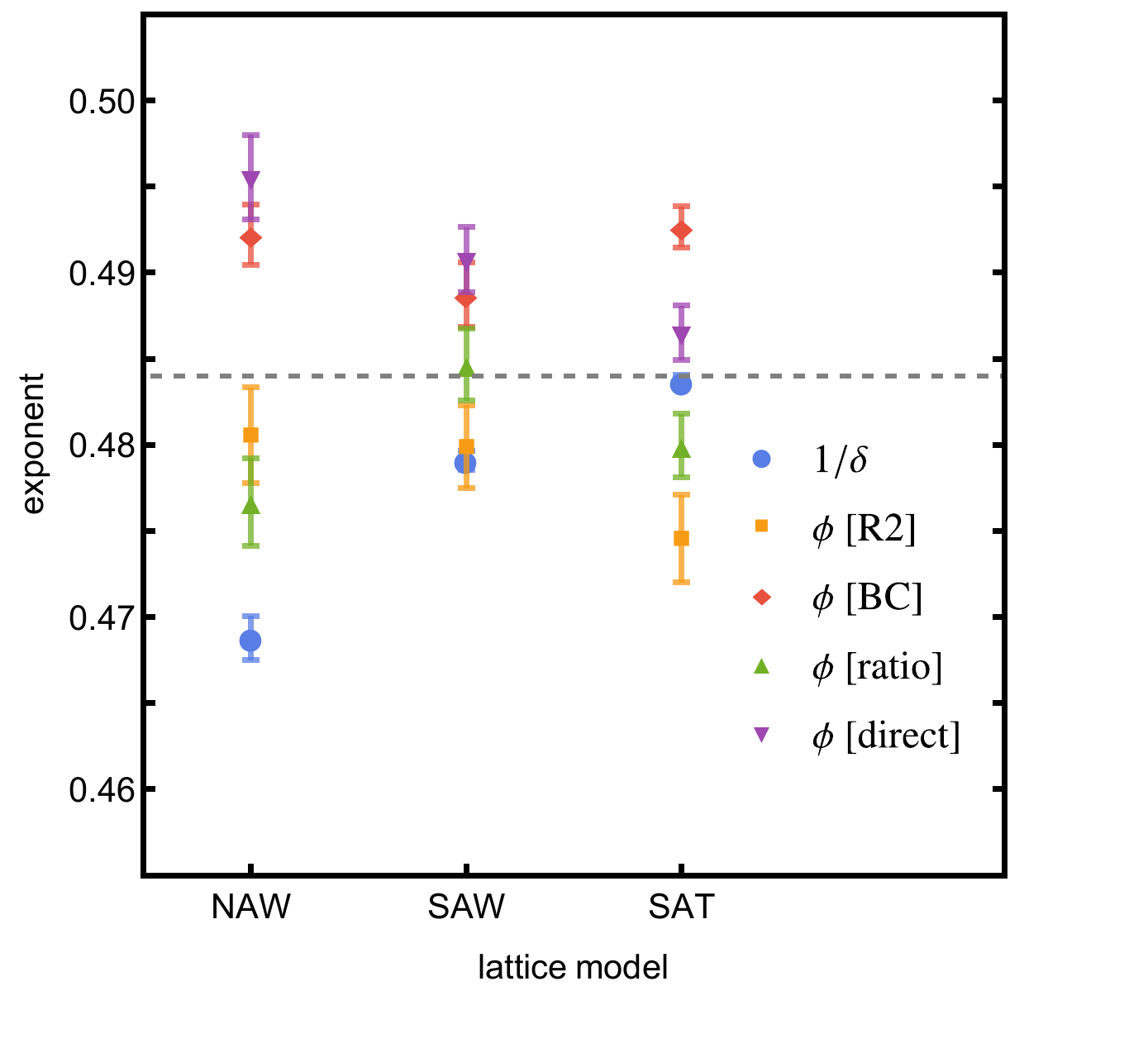}
	%\vspace{-0.5cm}
	\caption{Exponents for each lattice model. The dashed grey line marks the average estimate of the 3D crossover exponent $\phi=0.484(7)$. Data for SAW and SAT models are from \cite{Bradly2018}.}
	\label{fig:3DExponents}
	\vspace{-0.5cm}
\end{figure}

Following the method for SAWs and SATs outlined in \cite{Bradly2018} we then use the critical temperatures and other methods to calculate the exponents.
The results are shown in \fref{fig:3DExponents} and listed in \tref{tab:ExponentResults} alongside those for SAWs and SATs.
We stress that the error bars are due to statistical error, primarily from the curve fitting for the methods in \sref{sec:Scaling} and to a lesser extend from the simulation data.
Since each method is physically motivated based on known properties of the critical point, it is the spread in values that provide a complete estimation of the exponent $\phi$.
This spread is larger for NAWs than the other lattice models but not significantly so.
Despite the issues just mentioned, no single method stands out as better than the others.
Nor are we able to infer a trend between results for NAWs, SAWs and SATs.
If we understand that NAWs are infinitely repulsive SAWs and that SAWs are infinitely repulsive SATs then there is no clear indication that $\phi$ is not universal with respect to repulsive monomer-monomer interactions.

Finally, we can combine the results for all three lattice models for an estimate of the exponent for three dimensions.
This is done in a few steps.
First, we average the estimates of $\phi$ for the R2, BC and ratio methods together because while they are different ways to estimate the critical temperatures $\Tcn$, they all use the finite-size scaling form \eref{eq:UnScaling} to determine the exponent.
This makes them separate from the direct estimation of $\phi$ from \eref{eq:PhiRatio} and the $1/\delta$ exponent.
Then for each lattice model these three estimations of the critical exponent are averaged, under the assumption that \smash{$\phi=1/\delta$}.
\tref{tab:ExponentResultsBest} lists, for each lattice model, the critical temperatures $\Tcinf$, the value of $\phi$ averaged over only the finite-size scaling methods, and the value of $\phi$ using all methods.
Within error bars, all three lattice models agree on the value of $\phi$, and we report a value \smash{$\phi=0.484(7)$} for adsorption in three dimensions.

%%%%%%%%%%%%%%%%%%%%%%%%%%%%%%%%%%%%%%%%%%%%%%%%%%%%%%%%%%%%%%
\setlength{\tabcolsep}{2pt}
\begin{table}[t!]
	\caption{Exponents and critical temperatures for NAWs for each method. All values are from fits with correction-to-scaling terms.}
	\begin{tabular}{llll}
	\hline \hline
	 Method & \multicolumn{1}{c}{$1/\delta$} & \multicolumn{1}{c}{$\Tcinf$} & \multicolumn{1}{c}{$\phi$}  \\
	\hline
	
	$\Gamma$ 	& $0.4688(13)$ & $3.7557(85)$ & $-$ 	\\ 
	BC 		& $-$ 	& $3.707(12)$ 	& $0.4922(17)$ 		\\
	R2 		& $-$ 	& $3.7294(53)$ 	& $0.4805(28)$ 		\\
	ratio 	& $-$ 	& $3.726(11)$ 	& $0.4767(25)$ 		\\
	direct 	& $-$ 	& $-$ 			& $0.4955(25)$ 		\\
	
	\hline \hline
	\end{tabular}
% \vspace{-0.5cm}
	\label{tab:ExponentResults}
\end{table}
%%%%%%%%%%%%%%%%%%%%%%%%%%%%%%%%%%%%%%%%%%%%%%%%%%%%%%%%%%%%%%

%%%%%%%%%%%%%%%%%%%%%%%%%%%%%%%%%%%%%%%%%%%%%%%%%%%%%%%%%%%%%%
%%%%%%%%%%%%%%%%%%%%%%%%%%%%%%%%%%%%%%%%%%%%%%%%%%%%%%%%%%%%%%
\section{Conclusion}
\label{sec:Conc}

%%%%%%%%%%%%%%%%%%%%%%%%%%%%%%%%%%%%%%%%%%%%%%%%%%%%%%%%%%%%%%
\begin{table}[t]
	\caption{Best results for the adsorption temperature and the finite-size scaling estimates of $\phi$ for each lattice model. Bold values are the combined result for the crossover exponent for each lattice model and dimension. Data for SAW and SAT models are from \cite{Bradly2018}.}	
	\begin{tabular}{l|ll|l}
	\hline \hline
	 & \multicolumn{1}{c}{$T_c$} %& \multicolumn{1}{c}{$1/\delta$} 
	& \multicolumn{1}{c}{FSS $\phi$} & \multicolumn{1}{c}{$\phi$ [$=1/\delta$]}\\\hline	
	NAW		& 3.274(9) 	& 0.483(8)	& {\bf 0.482(13)} \\
	SAW		& 3.520(6) 	& 0.484(4)	& {\bf 0.485(6)} \\
	SAT		& 3.720(12)	& 0.482(9) 	& {\bf 0.484(2)} \\
	3D 			& 			& 			& {\bf 0.484(7)} \\
	\hline\hline
	\end{tabular}
	
	\label{tab:ExponentResultsBest}
\end{table}
%%%%%%%%%%%%%%%%%%%%%%%%%%%%%%%%%%%%%%%%%%%%%%%%%%%%%%%%%%%%%%

We have simulated the adsorption of neighbor-avoiding walks on the simple cubic lattice up to length 1024 using the flatPERM algorithm.
In addition to the further study of NAWs in their own right, the resulting critical behavior is in line with expectations from finite-size scaling theory.
We estimate the critical exponent for the adsorption transition to be \smash{$\phi=0.482(13)$} and the transition temperature is \smash{$\Tcinf=3.274(9)$}.
At first glance this value appears to be slightly lower than for SAWs or SATs but consistent within the error bar.
As was the case in our previous work \cite{Bradly2018}, what is not apparent in this final result is that there is a significant systematic error in any individual estimate of $\phi$ due to the variety of valid methods of analysis. 
This error is greater than the statistical error reported and has been shown to be consistent with analogous studies for the two dimensional case where the value of the critical exponent is not in doubt.
With that in mind, we conclude that the value of $\phi$ is in agreement with our previous values for SAWs and SATs on the simple cubic lattice.
Taken together, we have a value of \smash{$\phi=0.484(7)$} for three dimensions, which is in good agreement with other estimates \cite{Grassberger2005,Plascak2017}.
It also further confirms that $\phi$ deviates from the mean-field value of $1/2$ and is not super-universal.

In addition, the variety of methods includes a direct estimate of the $1/\delta$ exponent and as was the case for SAWs and SATs we find that it does not deviate from other methods of analysis as a means of estimating the critical exponent $\phi$.
This is further evidence against the case that the shift and crossover exponents are different in three dimensions.

%%%%%%%%%%%%%%%%%%%%%%%%%%%%%%%%%%%%%%%%%%%%%%%%%%%%%%%%%%%%%%
%%%%%%%%%%%%%%%%%%%%%%%%%%%%%%%%%%%%%%%%%%%%%%%%%%%%%%%%%%%%%%

Considering NAWs as a different class of lattice paths, the smaller critical temperature in comparison to SAWs follows from a reduction in entropy due to the stronger restriction on the allowed configurations.
Similarly, the critical temperature for SAWs is lower than that of SATs.
Considering NAWs as the infinitely repulsive limit of ISAW, $\phi$ has the same value as the non-interacting regime.
The same can be said for SAWs as the infinitely repulsive limit of ISAT, so we do not see any evidence that the critical exponent $\phi$ is not universal with respect to monomer-monomer interactions.
This is not a surprising result when considering that we are really modeling solvent quality and so the only effect of considering NAWs as opposed to SAWs is an enhanced effective excluded volume.

%%%%%%%%%%%%%%%%%%%%%%%%%%%%%%%%%%%%%%%%%%%%%%%%%%%%%%%%%%%%%
\begin{acknowledgments}
Financial support from the Australian Research
Council via its Discovery Projects scheme (DP160103562)
is gratefully acknowledged by the authors. C.~Bradly thanks the School of Mathematical Sciences, Queen Mary University of London for hospitality. Numerical simulations were performed using the HPC cluster at University of Melbourne (2017) Spartan HPC-Cloud Hybrid facilities at the University of Melbourne.
\end{acknowledgments}

%\bibliography{../polymers}{}

\end{document}